\documentclass[letterpaper,11pt]{article}
\usepackage{jheppub}

\usepackage{amsmath}
\usepackage{amsfonts}
\usepackage{amssymb}
\usepackage{bm}

\usepackage{mathrsfs}

\usepackage{hyperref}

\usepackage{comment}

\usepackage[dvipsnames]{xcolor}

\newcommand{\Z}{\mathbb{Z}}

\DeclareMathOperator{\tr}{tr}
\DeclareMathOperator{\tor}{Tor}
\DeclareMathOperator{\free}{Free}

\newcommand{\zero}{\text{zero}}

\title{Electromagnetic duality and entanglement anomalies}

\author[a]{William Donnelly}
\emailAdd{williamdonnelly@gmail.com}
\affiliation[a]{Department of Physics, University of California, Santa Barbara, CA 93106, USA}
\author[a]{Ben Michel}
\emailAdd{benjamin.l.michel@gmail.com}
\author[b]{Aron C. Wall}
\emailAdd{aroncwall@gmail.com}
\affiliation[b]{School of Natural Sciences, Institute for Advanced Study, Princeton, New Jersey 08540, USA}

\abstract{Duality is an indispensable tool for describing the strong-coupling dynamics of gauge theories. However, its actual realization is often quite subtle: quantities such as the partition function can transform covariantly, with degrees of freedom rearranged in a nonlocal fashion.  We study this phenomenon in the context of the electromagnetic duality of abelian $p$-forms.  A careful calculation of the duality anomaly on an arbitrary $D$-dimensional manifold shows that the effective actions agree exactly in odd $D$, while in even $D$ they differ by a term proportional to the Euler number. Despite this anomaly, the trace of the stress tensor agrees between the dual theories.
We also compute the change in the vacuum entanglement entropy under duality, relating this entanglement anomaly to the duality of an ``edge mode'' theory in two fewer dimensions.
Previous work on this subject has led to conflicting results; we explain and resolve these discrepancies.}

\begin{document}

\maketitle

\newpage

\section{Introduction}
Electromagnetic duality is a symmetry of many gauge theories, but the dual degrees of freedom are not local in the original variables. Thus one might expect probes of the localization of correlations -- such as entanglement entropy -- to depend on the duality frame.

One particularly tractable example is the electromagnetic duality of Maxwell theory, where the the field strength $F_{\mu \nu}$ is interchanged with its Hodge dual $(\star F)_{\mu \nu} = \frac12 \epsilon_{\mu \nu \rho \sigma}F^{\rho \sigma}$.\footnote{Note that we assume oriented manifolds throughout, since Hodge and Poincar{\'e} duality hold only in that case. 
Both of these dualities can be generalized to the non-orientable case \cite{Metlitski:2015yqa}, but we leave such a generalization to future work.}  
Depending on the context, this is also called Poincar{\'e} or $S$-duality, and generalizes to Maxwell theories with $p$-form potentials in $D$ spacetime dimensions.\footnote{These $p$-forms appear in string theory as the gauge fields coupled to D-branes \cite{Polchinski:1995mt}.}
The dual $\tilde{p}=D-p-2$-form potentials satisfy $dA = F$ and $d\tilde{A} = \star F$ and so determining one in terms of the other requires the nonlocal inversion of a differential operator.

For many purposes the dual theories are identical, \textit{e.g.}~the space of classical solutions is the same.
But on the quantum level there are conflicting opinions \cite{Duff:1980qv,Siegel:1980ax,Fradkin:1984ai,Grisaru:1984vk,Schwarz1984,Witten1995,Larsen:2015aia} about the extent to which the dual theories may be regarded as equivalent, and even whether the trace anomaly agrees when $p \ne \tilde{p}$. Quantum equivalence requires the dual partition functions $Z$ and $\tilde{Z}$ to be exactly equal, which is not always the case.  
For example the partition function of Maxwell theory in $D=4$ transforms under electromagnetic duality as a modular form \cite{Witten1995}, which characterizes the anomaly in the duality symmetry.

The results \cite{Schwarz1984,Witten1995} are in conflict. In \cite{Schwarz1984} Schwarz and Tyupkin compute the ratio of partition functions of dual $p$-form theories by computing a ratio of functional determinants, which come from Gaussian integrals over the non-zero modes of the Laplacian. This calculation yields a vanishing anomaly in even dimensions, but Witten later computed a nontrivial anomaly in ordinary 1-form Maxwell theory in four dimensions \cite{Witten1995}, which was confirmed by subsequent calculations \cite{McLellan:2012zy,Metlitski:2015yqa}. To our knowledge this conflict has not been resolved in the literature. However, our interest in the duality properties of entropy led to a more thorough calculation of the duality anomaly of $p$-form gauge theories, which enables us to reconcile these results. We find that previously-neglected zero modes and instantons contribute factors that 1) give rise to the even-dimensional anomaly and 2) trivialize the duality in odd dimensions. This proves, for example, quantum equivalence of the scalar and photon in $D=3$, and reproduces the known anomaly in $D=4$. Our method extends to arbitrary $D$ and $p$.\footnote{We do not consider the case of massive $p$-form theories, for which the duality relation is instead $\tilde{p}=D-p-1$, but \cite{Buchbinder:2008jf} argues that there is no duality anomaly in this case.}

In general, we find the following for the duality anomaly of an abelian $p$-form gauge theory on $M$:
\begin{equation}
\label{eq:dualityresult}
\log\frac{Z_p}{\tilde{Z}_{\tilde{p}}} =  \left\{
	\begin{array}{ll}
		(-1)^{p+1}\ \chi(M) \log\sqrt{\frac{q}{\tilde{q}}}  &\quad \quad D\mbox{ even} \\
		0 & \quad \quad D\mbox{ odd,}
	\end{array}
	\right. 
\end{equation}
where $Z_p$ is the partition function of the gauge theory and $\tilde{Z}_{\tilde{p}}$ the partition function of its electromagnetic dual. $q$ and $\tilde{q} = 2\pi / q$ are the couplings of the dual theories, which enter into the path integral of a U(1) gauge field via the flux quantization condition. The argument of the log is in units of a mass scale $\mu$ that must be introduced to define the quantum theory. The scale $\mu$ does not enter into the classical theory, the classical duality, or the quantum correlation functions, but it does enter into the anomalous quantum duality, where its role is to fix the units.  

The vanishing of the odd-dimensional anomaly follows from a theorem of Cheeger \cite{Cheeger1979} that equates the ratio of analytic torsion \cite{Ray1973} (a product of functional determinants related to the partition function of Chern-Simons theory) and Reidemeister torsion \cite{Reidemeister1935} (a combinatorial quantity invented in the 1930s to distinguish lens spaces) with a ratio of the sizes of torsion subgroups. 
Although this combination of quantities from entirely different branches of mathematics may seem obscure, each quantity appears naturally in the ratio of partition functions. This physical application of the Cheeger-M\"uller theorem may be of interest apart from our study of the duality properties of entanglement entropy.

The even-dimensional duality anomaly is purely topological, and may also be absorbed into a local counterterm.  In computations involving renormalization, often one already needs the leeway to shift the action by a local counterterm, and the duality anomaly may simply be absorbed.  But it other contexts knowing the exact form of the anomaly is important.\footnote{Examples include: (i) nonrenormalization theorems, where a finite shift in a quantity might violate the theorem, or (ii) $p$-form fields that arise from Kaluza-Klein reduction, where it is important to preserve the local covariance of the higher dimensional theory \cite{Donnelly:2015hxa}.}

Our calculation of entanglement entropy relies on the replica trick \cite{Calabrese:2004eu, Calabrese:2009qy}, which expresses the entropy in terms of a partition function. Thus one might expect that entanglement inherits some of the duality structure. In detail, we calculate the entanglement entropy $S_A = -\tr\rho_A\log\rho_A$ of a region $A$ by first computing $\tr\rho_A^n$, analytically continuing to non-integer $n$, then using the identity $S_A~=\lim_{n\rightarrow 1}(1-~n\partial_n)~\log\tr\rho_A^n$ to obtain the entropy. Specializing to the vacuum and constructing powers of the vacuum reduced density matrix as euclidean path integrals, one finds $\tr\rho_A^n=Z(M_A^{(n)})$, the partition function of the theory on the ``replica manifold" $M_A^{(n)}$ of index $n$. Calculation of the entropy is reduced to the calculation of a replica partition function, and if one identifies the replica index with an inverse temperature the entanglement entropy is its thermodynamic entropy at $n=1$. 

However, eq.~\eqref{eq:dualityresult} implies that $Z(M_A^{(n)})$ can transform anomalously under electromagnetic duality, and so vacuum entanglement may also depend on the duality frame. We call this phenomenon an ``entanglement anomaly''.
\footnote{Previously entanglement has been shown to transform anomalously under other symmetries, e.g. under a Lorentz boost in theories with chiral anomalies \cite{Wall:2011kb,Castro:2014tta,Nishioka:2015uka,Iqbal:2015vka,Azeyanagi:2015uoa,Hughes:2015ora}.
In this work we find an analogous effect when the duality symmetry is anomalous.}
It is given by
\begin{equation}
\label{eq:anomalyeq}
\Delta S_A = (1-n\partial_n) \log\frac{Z(M_A^{(n)})}{\tilde{Z}(M_A^{(n)})}
\end{equation}
evaluated at $n=1$. The ratio can be computed using \eqref{eq:dualityresult}. For a $p$-form theory at coupling $q$, the change in entanglement entropy $A$ is
\begin{equation}
\label{eq:anomalyresult}
\Delta S_A =  \left\{
	\begin{array}{ll}
		(-1)^{p-1}\ \chi(\partial A) \log\sqrt{\frac{q}{\tilde{q}}}  &\quad \quad D\mbox{ even} \\
		0 & \quad \quad D\mbox{ odd,}
	\end{array}
	\right. 
\end{equation}
where $\chi(\partial A)$ is the Euler characteristic of the entangling surface. This ratio is the duality anomaly of a $(p-1)$-form edge mode theory on the entangling surface. Since the partition function of an abelian gauge theory on a replica manifold contains replica index-independent pieces that correspond to edge modes living on the entangling surface \cite{Donnelly:2014fua,Donnelly:2015hxa}, the entanglement anomaly arises naturally from this effect. This is discussed in \S~\ref{subsection:edgemodes}.
We also consider theories with a  $\theta$-term, see \S~\ref{subsection:thetaterm}.

While the constant term in the even-$D$ entanglement entropy changes under rescalings of the cutoff \cite{Calabrese:2004eu}, we show in \S~\ref{section:anomalies} that the constant term in the even-dimensional entanglement anomaly is actually unchanged under simultaneous transformation of the two theories. This is consistent with recent results in the condensed matter literature \cite{Zhang:2011jd,Kim2015} and we give a general derivation.

Now we outline the body of the paper.  In \S~\ref{section:partitionfunctions} we describe our calculation of the partition function of $p$-form gauge theory on an arbitrary manifold and outline our calculation of the ratio of electromagnetic dual partition functions $Z_p/\tilde{Z}_{\tilde{p}}$. We explain how to reconcile the conflicting results of \cite{Schwarz1984,Witten1995} and why the anomaly vanishes in odd~$D$.  In \S~\ref{section:trace}, we explain why the stress tensor is the same for the dual theories in even~$D$ (including the trace), in agreement with the arguments of \cite{Grisaru:1984vk}.  Finally, in \S~\ref{section:anomalies} we use the partition function to compute the entanglement anomaly. 
We show that thermal entropy is duality-invariant and address the question of universality under a change of regulator, and conclude by interpreting the entanglement anomaly physically as the duality anomaly of an edge mode theory living on the entangling surface.

Details of the duality calculations are left to appendices~\ref{appendix:duality} and~\ref{appendix:thetaterm}. In appendix~\ref{appendix:1dmaxwell} we work out a simple example that illustrates the importance of zero modes: Maxwell theory in one spacetime dimension, which has no states besides the vacuum. The oscillator partition function fails to reproduce the trivial canonical sum over states, unless accompanied by the zero mode contribution. 

Recent related work includes \cite{Radicevic:2016tlt}, where the author considers the interplay between entanglement and duality in discrete spin systems; \cite{Huang:2016bkp}, which discusses the conformal $p$-form theories; \cite{Delcamp:2016eya}, which develops the extended Hilbert space in the magnetic representation. 
\cite{Raj:2016zjp} carried out some explicit calculations of $p$-form partition functions on the sphere and confirms the existence of the anomaly in even but not odd dimensions. 
\cite{Agon:2013iva} made use of duality to relate the entanglement entropy of a Maxwell theory to a compact scalar in $2+1$ dimensions; our results justify their use of this duality, since we show that it is exact in odd dimensions.

\section{Partition functions and duality}
\label{section:partitionfunctions}

In this section we calculate the partition function of $p$-form Maxwell theory and the ratio between the partition function and its electromagnetic dual.

\subsection{The partition function of $p$-form gauge theory}

Consider $p$-form electrodynamics on a compact manifold $M$, with gauge potential $A$ and field strength $F = dA$. 
The Euclidean action is\footnote{This defines our convention for the Hodge dual $\star$.}
\begin{equation} \label{action}
I = \frac{1}{2 q^2} \int_M (\star F) \wedge F = \frac{1}{2 (p+1)! q^2} \int_M \sqrt{g} F^{\mu \nu \cdots} F_{\mu \nu \cdots}.
\end{equation}
When $p=1$ this reduces to the familiar Maxwell action $\frac{1}{4 q^2} \int \sqrt{g} F_{\mu\nu} F^{\mu\nu}$.
The constant $q$ is the coupling constant, the fundamental unit of charge. 

We will compute the partition function by generalizing the approach developed in Ref.~\cite{Donnelly:2013tia} to $p$-form theories. The partition function on $M$ is given by the Euclidean path integral
\begin{equation} \label{Z}
Z = \sum_\mathrm{bundles} \int \mathcal{D}[A/\mathcal{G}]\,e^{-I[A/\mathcal{G}]}.
\end{equation}
The path integral is over all equivalence classes of connections, which we denote $A/\mathcal{G}$. 
This includes
a sum over all gauge bundles (allowing for field strengths $F$ that cannot be globally expressed as $F=dA$) 
and over connections with vanishing curvature $F=0$ (zero modes). 

We first decompose the field strength $F$ into a piece that comes from a globally-defined $p$-form potential and a piece that does not:
\begin{equation} \label{Fmodes}
F = \mathscr{F} + dA .
\end{equation}
where $A$ is the $p$-form potential and $\mathscr{F}$ is the part of $F$ that cannot be written as $dA$.
The Bianchi identity implies $dF=0$, so $\mathscr{F}$ is an element of the $(p+1)^\text{st}$ cohomology group of $M$.
The Dirac quantization condition further restricts $\mathscr{F}$ to be an element of the integer cohomology,
\begin{equation}
\mathscr{F} \in 2 \pi H^{p+1}(M, \Z).
\end{equation}
The decomposition \eqref{Fmodes} only fixes $\mathscr{F}$ up to the addition of an exact form. We can fix this remaining freedom by choosing $\mathscr{F}$ to be harmonic, $\Delta_{p+1} \mathscr{F} = 0$.
This choice makes the decomposition \eqref{Fmodes} orthogonal, and as a result the action $I$ splits as a sum over $\mathscr{F}$ and an integral over $A$:
\begin{equation}
I = \frac{1}{2 q^2} \int_M \left[(\star \mathscr{F}) \wedge \mathscr{F} + (\star d A) \wedge d A \right].
\end{equation}
The sum over instantons therefore decouples from the remainder of the partition function; we will return to it after first considering the functional integral over the potential $A$.

To carry out the integration over the $p$-form potential $A$, we write the mode expansion
\begin{equation} \label{Amodes}
A = A_\zero +  \sum_{n} \alpha_n A_n,
\end{equation}
where $A_n$ are the nonzero modes of the $p$-form Laplacian, $\Delta_p A_n = \lambda_n A_n$, which are chosen to be orthonormal. 
$A_\zero$ is the zero mode satisfying $\Delta_p A_\zero = 0$.
We will first deal with the nonzero modes, then treat $A_\zero$ separately.

We must introduce Faddeev-Popov ghosts that cancel out the unphysical polarizations in order to carry out the gauge-invariant path integral \eqref{action}.\footnote{One does not usually consider ghosts in abelian theories, since they decouple from the physical polarizations and hence do not contribute to correlation functions. However, the ghosts still contribute to the partition function and therefore to the entropy.
}
In $p$-form gauge theory, these ghosts are $(p-1)$-forms with fermionic statistics.
However, these $(p-1)$-forms have their own $(p-2)$-form gauge symmetry: some of the gauge transformations are redundant, and the ghosts subtract too many degrees of freedom.
It is then necessary to add in further positive degrees of freedom via ghosts-for-ghosts \cite{Townsend1979,Siegel:1980jj,Schwarz1984}.
One must continue in this way, introducing $k$-form fields for all $k = 0, \ldots, p$ with alternatingly bosonic and fermionic statistics.
The number of ghosts increases as the form degree decreases, so we have one $p$-form gauge field, two $(p-1)$-form ghosts, three $(p-2)$-form ghosts-for-ghosts, etc.

Having introduced the ghosts, the action for the field $A$ is non-degenerate, $I = \frac{1}{2 q^2} \langle A, \Delta_p A \rangle$, so we can carry out the path integral as usual. 
The space of $p$-forms comes with a natural measure induced by the inner product on $p$-forms.
Because the modes $A_n$ are orthonormal, this measure can be expressed in terms of the coefficients $\alpha_n$ of the mode expansion as 
\begin{equation} \label{measure}
\mathcal{D} A =  \prod_{n} \frac{\mu \, d \alpha_n}{\sqrt{2 \pi} q}.
\end{equation}
The reason for our choice of overall multiplicative constant $ 1 / \sqrt{2 \pi} q$ will become clear in the course of the calculation. We also had to introduce a parameter $\mu$, with dimensions of mass, since the measure must be dimensionless. Most quantities are independent of $\mu$ but it is part of the definition of the theory; it will set the units in the duality.

The path integral over nonzero modes of $A$ is a Gaussian integral, and hence reduces to a functional determinant
\begin{equation}
\int \mathcal{D} A e^{- \frac{1}{2 q^2} \langle A, \Delta_p A \rangle} = \prod_{n} \int \frac{\mu \, d \alpha_n}{\sqrt{2 \pi} q} e^{-\frac{1}{2 q^2} \lambda_n \alpha_n^2} = \prod_n \left(\frac{\lambda_n}{\mu^2}\right)^{-1/2} = \det \left(\frac{\Delta_p}{\mu^2}\right)^{-1/2}.
\end{equation}
Carrying out the analogous integrals for the various ghost fields then leads to a string of determinants:
\begin{align}
\label{oscillatorz}
Z_\text{osc} &= \det (\tilde \Delta_p)^{-1/2} \det (\tilde\Delta_{p-1})^{+1} \cdots \det (\tilde\Delta_0)^{(-1)^{p+1} \frac{p+1}{2}}  \nonumber \\
&= \prod_{k=0}^p \det(\tilde\Delta_k)^{(-1)^{p+1-k} (p+1-k)/2}.
\end{align}
where $\tilde{\Delta}_k = \Delta_k/\mu^2$. The exponent reflects the proliferation of ghosts, and the sign reflects their alternatingly fermionic and bosonic character.

Next, there is the integral over flat connections $A_\zero$.
Let $H^p(M)$ denote the space of harmonic $p$-forms, a real vector space of dimension $b_p$ (the $p^\text{th}$ Betti number).
For flat connections the action vanishes, so the path integral for these modes simply computes the volume of the space of flat connections in the measure \eqref{measure}. 
This volume is finite because we identify $p$-form potentials under large gauge transformations, which are elements of the integer cohomology group $H^p(M,\Z)$.
As a discrete abelian group, it splits into a free part (given by $b_p$ copies of $\Z$) and a torsion part (a finite abelian group):
\begin{equation}
H^p(M,\Z) = \free H^p(M,\Z) \oplus \tor H^p(M,\Z) = \mathbb{Z}^{b_p} \oplus T^p.
\end{equation}
First we deal with the free part. The subspace $\free H^p(M,\Z)$ is simply the space of harmonic $p$-forms whose integrals around all noncontractible $p$-dimensional surfaces are integers.
As a group it is equal to $b_p$ copies of the additive group of the integers: $\free H^p(M,\Z) = \Z^{b_p}$.
Thus two $p$-form potentials $A$ and $A'$ are equivalent if the integral of $A - A'$ over any $p$-dimensional surface is a multiple of $2 \pi$. 
This is just a generalization of the Aharonov-Bohm phase to higher-form gauge theories.

We define for each $k$ a topological basis $\{ w_i \}_{i=1}^{b_k}$ of $\free H^k(M,\Z)$. 
The space of flat connections can then be parametrized as
\begin{equation}
A_\zero = \sum_{i = 1}^{b_p} \beta_i w_i,
\end{equation}
where $\beta_i = [0,2 \pi)$.
Written in the topological basis, the inner product on $p$-forms $\Gamma_p$ has components 
\begin{equation}
[\Gamma_p]_{ij} = \int_M (\star w_i) \wedge w_j.
\end{equation}
Integrating over the space of flat connections modulo large gauge transformations with the measure \eqref{measure} then gives a factor of
\begin{equation}
\det \left( \frac{2 \pi\mu^2}{q^2} \Gamma_p \right)^{1/2}
\end{equation}
There is a similar term that appears for the $(p-1)$ form gauge transformations.
In dividing by the volume of the gauge group, we must also divide by zero modes of the gauge transformations, which are flat $(p-1)$-forms.
These gauge symmetries have their own gauge redundancy given by $(p-2)$-forms, we must \emph{multiply} by the volume of the space of flat $(p-2)$ forms.
Continuing in this way we obtain the complete zero mode contribution
\begin{equation}
\label{zeromodez}
Z_\text{zero} = \det \left( \frac{2 \pi\mu^2}{q^2} \Gamma_p \right)^{1/2} \det \left( \frac{2 \pi\mu^4}{q^2} \Gamma_{p-1} \right)^{-1/2} \cdots \det \left( \frac{2 \pi\mu^{2(p+1)}}{q^2} \Gamma_0 \right)^{(-1)^p / 2}.
\end{equation}
Note that for a manifold with a single connected component of volume $V$, $\Gamma_0$ is the $1 \times 1$ matrix $V$.
This generalizes the volume correction that appears in the path integral for Maxwell theory in refs.~\cite{Donnelly:2013tia,Metlitski:2015yqa} and is essential for unitarity.

The torsion part $T^k := \tor H^k(M,\Z)$ of the cohomology groups is perhaps less familiar, but will play an important role in the duality in the most general case. When we integrate over the space of flat $p$-forms we must divide by the large gauge transformations. Taking the quotient by the discrete subgroup $T^p$ simply amounts to dividing the partition function by the number of elements $|T^p|$. To account for torsion-valued $p-1$-form large gauge transformations we must multiply by $|T^{p-1}|$ and so on, resulting in the contribution
\begin{equation}
\label{ztors}
Z_\text{tors} = |T^p|^{-1} |T^{p-1}| \cdots |T^0|^{(-1)^{p+1}}
\end{equation}
to the partition function.

Finally we return to the instanton contribution, i.e. gauge connections that cannot be expressed as $F=dA$. The Bianchi identity identifies them as elements of the cohomology group
\begin{equation}
\mathscr{F} \in 2\pi \, H^{p+1}(M,\Z).
\end{equation}
Again, we can split the cohomology group into its free and torsion parts:
\begin{equation}
H^{p+1}(M,\Z) = \free H^{p+1}(M,\Z) \oplus \tor H^{p+1}(M,\Z) = \mathbb{Z}^{b_{p+1}} \oplus T^{p+1}.
\end{equation}
Elements of $\free H^{p+1}(M,\Z)$ are equivalence classes of $(p+1)$-forms, from which we choose $\mathscr{F}$ to be the unique harmonic representative.
Elements of the torsion subgroup are associated with vanishing field strengths, and hence these instantons simply lead to an overall factor of $|T_{p+1}|$.  (These correspond to flat connections that have nontrivial holonomy around certain noncontractible $p$-surfaces, and yet do not come from harmonic $p$-forms.)  Thus the full sum over instantons is given by
\begin{equation} \label{bundlez}
Z_\text{inst} = |T^{p+1}| \sum_{\mathscr{F} \in \mathbb{Z}^{b_{p+1}}} e^{-I[\mathscr{F}]}.
\end{equation}

Summarizing, we find that the partition function of $p$-form gauge theory is
\begin{equation}
\label{eq:partitionfunction}
Z_p = \prod_{k=0}^{p} \left( \frac{\det \left( \frac{2 \pi\mu^{2(p-k+1)}}{q^2} \Gamma_k \right)^{1/2} }{ |T^k| \det(\tilde\Delta_k)^{(p-k+1)/2} } \right)^{(-1)^{p-k}}
|T^{p+1}| \sum_{\mathscr{F} \in \mathbb{Z}^{b_{p+1}}} e^{-I[\mathscr{F}]}.
\end{equation}
Except for an anomaly in even dimensions, the factors of $\mu$ cancel between numerator and denominator. This anomaly fixes the units in \eqref{eq:dualityresult} but otherwise plays no role. We will not keep the factors of $\mu$ explicit in the remainder of this section, deferring the details to the end of appendix~\ref{appendix:duality}.

Next we compute the change in the effective action under electromagnetic duality.

\subsection{Electromagnetic duality}

With the expression \eqref{eq:partitionfunction} it is a straightforward exercise to compute the relation between the dual partition functions. We will use zeta function regularization. The partition function of the electromagnetic dual is \eqref{action} with $p\rightarrow \tilde{p}=D-p-2$ and $q\rightarrow \tilde{q} = 2\pi/q$, which is equivalent to the replacement
\begin{equation}
\tilde{F} = \left(\frac{\tilde{q}}{q}\right)\star{F}.
\end{equation}
Poincar\'e duality and Poisson summation are the only tools needed to compute the ratio $Z_p / \tilde{Z}_{\tilde{p}}$, a task we defer to appendix~\ref{appendix:duality}.

First we isolate the oscillator contribution to the ratio. The ratio of oscillator partition functions is
\begin{equation}
\frac{Z_\text{osc}}{\tilde{Z}_\text{osc}} = \left[\prod_{k=0}^D \left(\det \Delta_k\right)^{k(-1)^{k+1}/2}\right]^{(-1)^{p+1}}.
\end{equation}
This expression is related to a quantity known as the Ray-Singer analytic torsion \cite{Ray1971,Ray1973},
\begin{equation}
\label{eq:RS}
\tau_\text{RS} = \prod_{k=0}^D \left(\det \Delta_k\right)^{k(-1)^{k+1}/2}.
\end{equation}
It plays a role in abelian Chern-Simons theory as the magnitude squared of the partition function \cite{McLellan:2012zy} but was originally defined as an analytic analog to a combinatorial invariant called Reidemeister torsion, which we will encounter soon.

When $D$ is even, $\tau_\text{RS}= 1$ by Poincar\'e duality and so the ratio of oscillator partition functions is
\begin{equation}
\log\frac{Z_\text{osc}}{\tilde{Z}_{\text{osc}}} =
\left\{
	\begin{array}{ll}
		0  &\quad \quad D =2n \\
		(-1)^{p+1}\ \log \tau_\text{RS} & \quad \quad D=2n+1.
	\end{array}
	\right.
\end{equation}
This is the result obtained by Schwarz and Tyupkin \cite{Schwarz1984}, who considered only the oscillator modes. Note in particular that the contribution to the anomaly vanishes in even $D$.

The rest of $Z_p / \tilde{Z}_{\tilde{p}}$ comes from the zero modes and instantons. Making the simplifying assumption that the torsion subgroups of $H^k(M,\mathbb{Z})$ are trivial, they contribute
\begin{align}
\frac{Z_\text{zero} Z_\text{inst}}{\tilde{Z}_\text{zero}\tilde{Z}_\text{inst}} &=\left[ \left(\frac{2\pi}{q^2}\right)^{\chi}\prod_{k=0}^{D} \det\left(\Gamma_k\right)^{(-1)^{k}}\right]^{(-1)^p/2}
\end{align}
which follows from Poincar\'{e} duality and the relation $\chi=\sum (-1)^k b_k$ between the Euler characteristic and the Betti numbers.
This ratio is related to the Reidemeister torsion:
\begin{equation}
\tau_\text{Reid} = \prod_{k=0}^D \det \Gamma_k^{(-1)^{k}/2}.
\end{equation}
This flavor of torsion was invented in 1935 to classify lens spaces \cite{Reidemeister1935}, which have the same homotopy groups but are not homeomorphic. It is actually the oldest non-homotopy invariant \cite{Turaev1986}, and is defined in terms of chain complexes on $M$. Ray and Singer defined their torsion \eqref{eq:RS} as an analytic analog in the early 1970s. Cheeger and M\"uller independently proved a few years later that the two are actually equal (up to the torsion subgroups $T^k$ to be discussed), culminating in the Cheeger-M\"uller theorem \cite{Cheeger1977,Cheeger1979,Muller1978}
\begin{equation}
\label{eq:cheegermuller}
    \tau_\text{RS} = \tau_\text{Reid}.
\end{equation}
This equation is the key to the triviality of the odd-dimensional anomaly. Thus
\begin{equation}
\frac{Z_\text{zero} Z_\text{inst}}{\tilde{Z}_\text{zero}\tilde{Z}_\text{inst}} =
\left\{
	\begin{array}{ll}
		(-1)^{p+1}\ \chi(M) \log\sqrt{\frac{q}{\tilde{q}}}  &\quad \quad D =2n \\
		(-1)^{p}\ \log \tau_\text{Reid} & \quad \quad D=2n+1,
	\end{array}
	\right.
\end{equation}
and the duality anomaly of a $p$-form theory on a $D$-manifold $M$ is
\begin{align}
\label{eq:dualityresult2}
\log \frac{Z_p} {\tilde{Z}_{\tilde{p}}} &=
\left\{
	\begin{array}{ll}
	    (-1)^{p+1}\ \chi(M) \log\sqrt{\frac{q}{\tilde{q}}}  &\quad \quad D=2n \\
		(-1)^{p+1}\log\frac{\tau_\text{RS}}{\tau_\text{Reid}}=0, & \quad \quad D=2n+1.
	\end{array}
\right.
\end{align}
In odd $D$ the zero mode and instantons cancel the oscillator contribution to the ratio and so the duality is exact. Our even-dimensional result agrees with Witten's computation \cite{Witten1995} of the duality anomaly of $D=4$ Maxwell theory; in appendix~\ref{appendix:thetaterm} we reproduce the $\theta$-dependence and discuss a phase. Only the zero modes and instantons contribute to the anomaly, which is why Schwarz and Tyupkin found an exact duality. It is somewhat ironic that the zero modes and instantons trivialize the odd-dimensional duality instead.

The quantity $\frac{q}{\tilde q}$ in \eqref{eq:dualityresult2} has mass dimension $2(p+1)-D$ and so must be accompanied by a dimensionful factor; this is furnished by the  parameter $\mu$ that we had to introduce in order to make the measure \eqref{measure} dimensionless. We will see in appendix~\ref{appendix:duality} that an anomaly in rescaling $\mu$ out of the functional determinants multiplies \eqref{eq:dualityresult2} by an extra term
\begin{align}\label{eq:mus}
(-1)^p\chi(M) \log\mu^{p+1-D/2}, 
\end{align}
after application of the McKean-Singer formula \cite{McKean1967}. This precisely fixes the units.

Last we consider the contribution from any torsion subgroups $T^k\subset H^k(M,\mathbb{Z})$. In the presence of nontrivial $T^k$, Cheeger \cite{Cheeger1979} found that the relation \eqref{eq:cheegermuller} is modified to
\begin{equation}
\label{nontors}
\frac{\tau_\text{RS}}{\tau_\text{Reid}}=\prod_{k=0}^D |T^k|^{(-1)^{k+1}}
\end{equation}
and so to maintain duality invariance we must show that their effect on the ratio of partition functions is to divide by this factor. In appendix~\ref{appendix:duality} we show that their contribution to the ratio of partition functions is
\begin{align}
\frac{Z_{p,\text{tors}}}{\tilde{Z}_{\tilde{p},\text{tors}}} &= \left[ \prod_{k=0}^D |T^k|^{(-1)^{k+1}}\right]^{(-1)^p}
\end{align}
which is trivial in even dimensions and cancels the non-torsion contribution in odd dimensions. In fact, the vanishing of the odd-dimensional anomaly is equivalent to Cheeger's refinement of the Cheeger-M\"uller theorem.

We find it rather surprising to find any application to physics in this somewhat obscure relation between quantities from different branches of mathematics.

\section{Trace of the stress tensor}\label{section:trace}

Since the effective action in even dimensions is not invariant under a duality transformation, it is natural to wonder whether the duality shifts the value of any other observables.  One natural choice of observable is the stress tensor, obtained by varying the effective action with respect to the metric:
\begin{equation}
T_{ab} = -2 \frac{\delta}{\delta g^{ab}} \log Z
\end{equation}

In our regulator scheme, it is easy to see that $T_{ab}$ will \emph{not} be affected by a duality transformation: the duality anomaly is simply a finite number times a topological invariant $\chi$, which is independent of the metric.  Therefore, $T_{ab}$ is the same in both theories.

Nevertheless, there has been a considerable amount of confusion about this topic in the literature, due to an apparent discrepancy in the trace anomaly between the dual theories. In order to clarify this issue, we first remind the reader that we needed three dimensionful parameters in order to define the quantum partition function \eqref{Z} (apart from any geometric parameters of $M$):
\begin{description}
    \item ${\bm \mu}$: the dimensional measure factor appearing in each mode of the path integral;
    \item ${\bm \Lambda}$: an additional parameter present in some regularization schemes (such as an ultraviolet momentum cutoff or lattice scale);
    \item ${\bm q}$: the fundamental charge, which is dimensionful when $D \ne 2p + 2$.
\end{description}
The first two quantities, $\Lambda$ and $\mu$, do not appear in the classical theory but are needed to make the quantum theory well-defined.  

Note that some regularization schemes, such as the zeta-function regularization we have been using, 
introduce only a single dimensionful scale and are effectively identifying $\Lambda = \mu$.
In this case even a single mode in the partition function gives rise to a logarithmic divergence as $\Lambda \to \infty$.  
While consistent, this can lead to confusing outcomes like apparent UV divergences in theories with no local degrees of freedom, such as $D = 2$ Maxwell theory \cite{Donnelly:2012st}.  In such a scheme, our duality anomaly will \emph{appear} to take the form of a logarithmic divergence proportional to $\chi$, although it really comes from the zero mode integrals.

This point is important because the trace anomaly is frequently calculated using the log divergences of the theory.  However, it is necessary to consider the dependence on the logs of all three kinds of dimensionful parameters to get the correct result. Thus, suppose we have a compact manifold $M(R)$ with ``radius'' $R$, where $M(R)$ is given by acting on $M(1)$ with a uniform scaling factor $\Omega = R$.  Then the dependence of the effective action on $\log R$ is given by
\begin{equation}\label{eq:globalT}
\frac{\partial}{\partial \log R} \log Z_p = -\int_M T.
\end{equation}
Dimensional analysis now says that this term can be calculated if you know how $\log Z$ scales when you simultaneously adjust the mass scales of $\Lambda$, $\mu$, and $q$:
\begin{equation}
\frac{\partial}{\partial \log R} \log Z_p =
\left[\frac{\partial}{\partial \log \Lambda} + 
\frac{\partial}{\partial \log \mu} + 
\right(p+1-\frac{D}{2}\left) \frac{\partial}{\partial \log q} \right] \log Z_p
\end{equation}

The simplest case is that of a conformal $p$-form field in $D = 2p + 2$ dimensions, where the theory is dual to another $p$-form of the same rank.  In this case, $q$ is dimensionless.  Since the theory is conformal, $T = 0$ classically, $q$ is dimensionless, and any nonzero value of $T$ must come from quantum anomalies.  After setting $\mu = \Lambda$, balancing of logarithms (i.e. demanding that their arguments be dimensionless) requires that any dependence on $\log \Lambda$ must match with the dependence of $\log Z$ on a local conformal rescaling $g_{ab}^\prime = \Omega^2 g_{ab}$ of the metric.  Hence the trace $T = T_{ab} g^{ab}$ may be calculated from the log divergences of the theory.  However, $p = \tilde{p}$, so the log divergences of the two theories are identical, and the duality anomaly \eqref{eq:dualityresult2} is merely a finite function of $q$. Note that $T$ is determined locally in this conformal case.

Things are more subtle when $D \ne 2p + 2$.  In this case, the action \eqref{action} is not conformal, even classically.  Thus, in general $T$ may depend in a complicated and nonlocal way on the state of the fields.  The classical theory is \emph{almost} invariant under the global scale-invariance ($\Omega = \text{constant}$) generated by \eqref{eq:globalT}, but even this symmetry is partially broken by flux quantization effects that depend on $q$.

Now we discuss electromagnetic duality. In even dimensions, \eqref{eq:dualityresult2} and \eqref{eq:mus} tell us that the duality anomaly coming from zero modes is proportional to 
\begin{equation}
\Delta \log Z \propto \log\left(\frac{q}{\mu^{p+1-D/2}}\right)
\end{equation}
This logarithm is already balanced, and hence it does not produce any $\log R$ dependence.  Therefore, the integrated trace is unaffected by the duality, consistent with our claim above.

This conclusion is essentially the same as that of \cite{Grisaru:1984vk}, who argued that the \emph{total} trace of the stress tensor is duality invariant, even though the amount attributable to varying $\log \Lambda$\footnote{which they confusingly refer to as the ``trace anomaly''} depends on the duality frame. However, they regulated their zero modes by inserting a small mass, while we consider a U(1) gauge field whose IR divergences are regulated by finite-$q$ effects.

In the next section we will consider the effects of the duality anomaly on the entanglement entropy $S$.  Unlike $T_{ab}$, $S$ is sensitive to topological terms, and thus we will find a nonzero shift under duality.

\section{Entanglement anomalies}
\label{section:anomalies}

We now derive the entanglement anomaly, i.e. the difference in entanglement entropies between the $p$-form theory and its dual.
We use the replica trick, which enables us to compute the difference of entropies using the results of \S~\ref{section:partitionfunctions} in conjunction with the definition of the entanglement anomaly \eqref{eq:anomalyeq}. 
The procedure is straightforward: we substitute the replica manifold $M=M_A^{(n)}$ into \eqref{eq:dualityresult2}, then determine the $n$-dependence in order to compute the entanglement anomaly \eqref{eq:anomalyresult}. 
As described in the previous section the anomaly vanishes in odd spacetime dimension.

\subsection{The anomaly}

The change in vacuum entanglement of a region $A$ under duality \eqref{eq:anomalyeq} depends on the ratio of replica partition functions, i.e. the duality anomaly \eqref{eq:dualityresult2} of the theory on the replica manifold $M_A^{(n)}$. This is determined by the Euler characteristic of $M_A^{(n)}$, which follows immediately from its cut-and-paste construction: $M_A^{(n)}$ consists of $n$ copies of $M\setminus A$, glued together along $n$ copies of $A\setminus \partial A$, all glued to a single copy of the entangling surface $\partial A$. Since each piece is disjoint their Euler characters just add:
\begin{align}
\chi(M_A^{(n)}) &= n \chi(M\setminus A) + n\chi(A\setminus \partial A) +\chi(\partial A)\nonumber\\
&= n\chi(M\setminus \partial A) + (1-n)\chi(\partial A).
\end{align}
Using
\begin{equation}
\log\frac{Z(M_A^{(n)})}{\tilde{Z}(M_A^{(n)})} = (-1)^{p+1}\ \chi\left(M_A^{(n)}\right) \log\sqrt{\frac{q}{\tilde{q}}}
\end{equation}
together with the thermodynamic expression for the anomaly \eqref{eq:anomalyeq}, we find that the change in the entanglement entropy of a region $A$ in $p$-form Maxwell theory under electromagnetic duality is
\begin{equation}
\label{eq:evendanomaly}
\Delta S_A = (-1)^{p+1}\ \chi(\partial A) \log\sqrt{\frac{q}{\tilde{q}}}.
\end{equation}
Here $\tilde{q}=2\pi/q$ is the coupling of the dual gauge theory. We will argue that the entanglement anomaly arises physically from the global anomaly of a $p-1$-form edge mode theory living on the entangling surface; see \S~\ref{subsection:edgemodes}.

Now we discuss the universality of our results. A constant term in the entanglement entropy in even dimensions can usually be absorbed into a shift of the cutoff, and so our results may appear to depend on the choice of renormalization scheme.  For simplicity consider the entanglement entropy of a ball-shaped region in $D=4$ flat vacuum, whose general form is \cite{Calabrese:2004eu}
\begin{equation}
S = c_1 R^2\Lambda^2 + c_2 \log(R\Lambda) + c_3
\end{equation}
where $R$ is the radius of the sphere, $\Lambda$ is the cutoff, and $c_i$ are various constants. In the absence of duality only $c_2$ is universal, as changes in $c_{1,3}$ can be absorbed into shifts of the cutoff. The entanglement entropy of the same region in the dual theory is
\begin{equation}
\tilde{S} = \tilde{c}_1 R^2\tilde{\Lambda}^2 + \tilde{c}_2 \log(R\tilde{\Lambda}) + \tilde{c}_3.
\end{equation}
Universality of the coefficient of the log guarantees $c_2 = \tilde{c}_2$.

In these calculations of entanglement, $\Lambda$ is a physical inverse distance to the entangling surface and should be matched between the two theories in order to compare like quantities. If we rescale $\Lambda\rightarrow \Lambda' = \alpha \Lambda$ in the original theory, $c_3$ picks up a shift proportionate to $c_2$. In the dual theory we must do the same rescaling and so $\tilde{c}_3$ picks up a shift, but universality of the log coefficient guarantees that it matches the shift in $c_3$.  Thus, the entanglement anomaly $\Delta c_3 = c_3 - \tilde{c}_3$ is independent of $\Lambda$.\footnote{We are grateful to Mark Srednicki for pointing out the preceding argument.}

More generally, we expect that $\Delta c_3$ does not depend on the choice of (reasonable) regulator scheme for the theory.  Recall that the entanglement anomaly arises purely from the zero modes and instantons of the theory, since (in even dimensions, where the anomaly can exist) the nonzero modes of the dual theories are in correspondence.  Therefore, so long as (a) we use the same regulator for modes of the same wavelength on both sides of the duality, and (b) the regulator only affects UV divergent quantities, not zero modes or instantons, it follows that $\Delta c_3$ is a universal quantity.

\subsection{With a $\theta$ term}
\label{subsection:thetaterm}
We can also calculate the entanglement anomaly in the presence of a topological term in the action. For concreteness consider the theory with $p=1$ and $D=4$, with action
\begin{equation}
I=\frac{2\pi^2}{q^2}\left(\frac{1}{8\pi^2}\int_M F_{\mu\nu}F^{\mu\nu}\right) + \frac{i\theta}{2}\left(\frac{1}{8\pi^2}\int_M \frac{1}{2} \epsilon_{\mu\nu\rho\sigma}F^{\mu\nu}F^{\rho\sigma}\right).
\end{equation}
We derive the duality relation in appendix~\ref{appendix:thetaterm}. Here we just quote the result:
\begin{align}
\label{eq:4dmodular}
\tilde{Z} &= Z\left(\frac{-1}{\tau}\right) = e^{i\pi\sigma/4} \bar{\tau}^{(\chi+\sigma)/4}\tau^{(\chi-\sigma)/4}Z(\tau).
\end{align}
where $\tau = \frac{\theta}{2 \pi} + \frac{2 \pi i}{q^2}$ and $\sigma = b_2^+ - b_2^-$ is the topological (as opposed to metric) signature of the manifold. The partition function transforms as a modular form up to a phase.

Before we can apply this result to the entanglement anomaly, we need to determine the signature of the replica manifold.
The contribution of the phase in \eqref{eq:4dmodular} to the entanglement anomaly is
\begin{equation}
\Delta S_\text{phase} = \left(-\frac{i\pi}{4}\right)(1-n\partial_n) \sigma(M_A^{(n)})|_{n=1}
\end{equation}
If this quantity is nonzero the phase would contribute an imaginary piece to the entanglement anomaly. 
Since entropy is a real quantity, this is only consistent if $\sigma(M_A^{(n)})$ is linear in $n$ so that it does not contribute to the anomaly.
It was shown in \cite{Fursaev:1995ef} that this is indeed the case, at least for $D=4$: the signature is well-defined for manifolds with conical defects, and is linear in the replica number $n$.
We conjecture that this will also be the case for general $D$.
While we do not have a complete proof, we note that the non-additivity of the signature is given by a result of \cite{Wall1969}, and appears to vanish by symmetry considerations.

Assuming that the signature contribution vanishes, we can now substitute $\chi(M_A^{(n)})$ and $\sigma(M_A^{(n)})$ into \eqref{eq:anomalyeq} to find the generalization of  \eqref{eq:evendanomaly}:
\begin{equation}
\Delta S_A = -\frac{1}{4} \log \left[ \left( \frac{\theta}{2 \pi} \right)^2 + \left(\frac{2\pi}{q^2}\right)^2 \right] \chi(\partial A).
\end{equation}
This analysis extends to the case where $D$ is a multiple of $4$ and $p+1=D/2$. A case not covered by this analysis is $D=2$ Maxwell theory with a $\theta \int F$ term; in appendix~\ref{appendix:thetaterm} we show that the entanglement anomaly is independent of such a $\theta$ term.

\subsection{Invariance of the thermal entropy}

The results of the previous sections imply that thermal entropy does not change under electromagnetic duality. This is reassuring, since the total number of degrees of freedom should be duality invariant.

We compute thermal entropies using ordinary thermodynamics. The first law of thermodynamics $S_\text{therm}=\beta\langle E\rangle + \log Z_\text{therm}$ relates thermal entropy to the thermal partition function $Z_\text{therm}=\tr e^{-\beta H}$. When the theory lives on a spatial manifold $\Sigma$, the thermal partition function is equal to the path integral on $M^{(\beta)}=\Sigma \times S^1_{\beta}$. We rewrite the first law as
\begin{equation}
S_\text{therm}(\beta) = (1-\beta \partial_\beta) \log Z(M^{(\beta)})
\end{equation}
and specialize again to the case of $p$-form theories. Under electromagnetic duality, the change in thermal entropy is
\begin{equation}
\Delta S_\text{therm}(\beta) = (1-\beta \partial_\beta) \log \frac{Z_p(M^{(\beta)})}{\tilde{Z}_{\tilde{p}} (M^{(\beta)})}.
\end{equation}
We can compute the right hand side using \eqref{eq:dualityresult2} with $M=M^{(\beta)}$. This gives
\begin{equation}
\log\frac{Z_\text{therm}}{\tilde{Z}_\text{therm}}=
\left\{
	\begin{array}{ll}
		(-1)^{p+1}\ \chi(M^{(\beta)}) \log\sqrt{\frac{q}{\tilde{q}}}  &\quad \quad D =2n \\
		0 & \quad \quad D=2n+1.
	\end{array}
\right.
\end{equation}
The Euler character of a product manifold $A\times B$ satisfies $\chi(A\times B) = \chi(A) \cdot \chi(B)$, so the Euler character of $M^{(\beta)} = \Sigma \times S^1_{\beta}$ vanishes and hence $\Delta S_\text{therm}(\beta)=0$ in even dimensions.

Thus for any $p$ and $D$
\begin{equation}
S_\text{therm} = \tilde{S}_\text{therm}.
\end{equation}

\subsection{Edge modes}
\label{subsection:edgemodes}

In theories with gauge symmetry there is a question of how to define the entanglement entropy.
For example, \cite{Casini:2013rba} proposed a number of inequivalent definitions in terms of the algebra of observables inside the entangling surface.
Here we have adopted the definition of entanglement entropy via the replica trick.
The replica trick requires no additional input, such as a choice of algebra, and so must single out a particular definition.
In Refs.~\cite{Donnelly:2014fua,Donnelly:2015hxa} it was shown that the replica trick coincides with the ``extended Hilbert space'' definition of entanglement entropy for 1-form gauge fields.
In the case of abelian gauge fields, this coincides with what Ref.~\cite{Casini:2013rba} calls the ``electric'' definition of entanglement entropy and in condensed matter is sometimes called the ``rough edge'' \cite{Kim2015}.

Analysis of these replica partition functions $Z(M_A^{(n)})$ \cite{Donnelly:2015hxa} yields a decomposition into a bulk and an edge piece:

\begin{equation}
Z(M_A^{(n)}) = Z_\text{bulk} Z_\text{edge}
\end{equation}
where $Z_\text{bulk}$ describes degrees of freedom on $M\setminus \partial A$ and $Z_\text{edge}$ describes degrees of freedom on the entangling surface $\partial A$.

The edge mode partition function is
\begin{equation}
\label{eq:zedge}
Z_\text{edge} = \int \mathcal{D} E_\perp  e^{-I_\text{cl}(E_\perp)};
\end{equation}
the sum is over configurations of the normal electric field at the entangling surface and the exponent is the action of a classical solution with corresponding boundary configuration --- the action is itself a boundary term.
The reduced density matrix thus splits into superselection sectors labelled by $E_\perp$: $\rho = \oplus p(E_\perp) \rho_{E_\perp}$. 
This gives rise to a Shannon term $-\sum p(E_\perp) \log p(E_\perp)$ in the entanglement entropy \cite{Donnelly:2011hn}.

In this case the entropy of the edge modes is given by the log of the partition function of a ghost scalar confined to the entangling surface:
\begin{equation}
S_\text{edge} = -\log Z_{p=0}(\partial A).
\end{equation}
This suggests that the generalization of the edge mode entropy for a $p$-form theory is the partition function of a ghost $(p-1)$-form theory on the codimension-2 entangling surface.

Note that when the $p$-form theory in $D$ dimensions is conformal, the $(p-1)$-form theory in $D-2$ dimensions is also conformal.
In this case there is a universal logarithmic divergence in the entanglement entropy of a sphere related to the conformal anomaly of the theory \cite{Solodukhin:2008dh,Casini:2011kv}.
For example, the sphere entropy of Maxwell theory in $D=4$, contains a contribution from a ghost scalar in $D=2$ which is necessary in order to obtain agreement with the conformal anomaly \cite{Donnelly:2014fua,Huang:2014pfa,Donnelly:2015hxa}.

The results of the present paper provide further evidence for this edge mode theory. 
The anomaly in the entropy \eqref{eq:evendanomaly} is precisely the duality anomaly of a ghost $(p-1)$-form confined to the entangling surface.
Thus the universal differences in the electric versus magnetic prescriptions for the entanglement entropy  appears to be captured in the electromagnetic duality anomaly of the edge mode theory.
It would be interesting to either confirm or refute this conjecture, for example by calculating the edge mode contribution to the logarithmic divergence of the sphere entanglement entropy in conformal $p$-form theories.  In doing so one ought to pay close attention to the zero modes of the edge system \cite{Donnelly:2014fua,Donnelly:2015hxa,Huang:2016bkp}.

\section{Discussion}

Duality is a rich subject and we have only explored simple, abelian examples.  It would be interesting to extend our results to other dualities. In most cases this will not be easy: generically one cannot compute the partition function, and even when one can, the replica manifold may break symmetries (such as supersymmetry) that enabled the computation in the first place. However, any tractable calculations of entanglement in dual theories would be amenable to an anomaly analysis akin to ours.

It would also be interesting to understand the lattice analogue of the phenomenon we have described. There is an ambiguity in how to cut up the lattice in calculations of entanglement: one can put the cut in the middle of a plaquette, or on a vertex, or on an edge. When such a theory enjoys an electromagnetic duality, its dual lives on the dual lattice and so inherits a different prescription for the entropy. In self-dual theories the entanglement anomaly should record this dependence on the choice of prescription.

\acknowledgments

We are very grateful to Max Metlitski for discussions of many kinds of torsion, to Mark Srednicki for pointing out the universal nature of the anomaly, and to Tarun Grover and Brayden Ware for discussions of the implications for the lattice. 
We would also like to acknowledge the hospitality of the Perimeter Institute during several phases of this project.
BM is supported by NSF Grant PHY13-16748.
AW is supported by the Institute for Advanced Study, the Martin A. and Helen Chooljian Membership Fund, the Raymond and Beverly Sackler Foundation, and NSF grant PHY-1314311.

\appendix

\section{Electromagnetic duality}
\label{appendix:duality}

In this appendix we derive \eqref{eq:dualityresult2}. For simplicity we first consider the case where the torsion subgroups of the cohomologies $H^k(M,\mathbb{Z})$ are trivial, also deferring discussion of the dimensionful factors $\mu$ appearing in the measure \eqref{measure} to the end.

As described in \S~\ref{section:partitionfunctions}, the partition function of $p$-form Maxwell theory on a manifold $M$ decomposes as 
\begin{equation}
Z_p = Z_\text{osc}Z_\text{zero}Z_\text{inst}
\end{equation}
where $Z_\text{osc}$, $Z_\text{zero}$ and $Z_\text{inst}$ can be found in \eqref{oscillatorz}, \eqref{zeromodez} and \eqref{bundlez}. In the dual $\tilde{p}$ form theory, the oscillator determinants $\tilde{Z}_\text{osc}$ and zero mode factors $\tilde{Z}_\text{zero}$ are simply obtained from \eqref{oscillatorz} and \eqref{zeromodez} by replacing $p$ with $\tilde{p} = D-p-2$.

Relating the instanton partition functions is slightly more involved. The trick is to do a Poisson summation and make use of  Poincar\'{e} duality. The instanton partition function of the original theory is
\begin{align}
\label{bundleduality}
Z_\text{inst} &= \sum_{ \mathscr{F}\in H^{p+1}(M,\mathbb{Z})} e^{-I_\text{cl}(\mathscr{F})} = \sum_{\vec{m}\in \mathbb{Z}^{b_{p+1}}} e^{- \frac{1}{2} \left(\frac{2 \pi}{q}\right)^2 \vec m \cdot  \Gamma_{p+1} \cdot \vec m}\nonumber\\
&= \det \left( \frac{2 \pi \Gamma_{p+1}}{q^2} \right)^{-1/2} \cdot \sum_{\vec{n}\in \mathbb{Z}^{b_{\tilde{p}+1}}} e^{- \frac{1}{2} \left(\frac{2 \pi}{\tilde{q}}\right)^2 \vec n \cdot  \Gamma_{\tilde{p}+1} \cdot \vec n}\nonumber\\
&= \det \left( \frac{2 \pi \Gamma_{p+1}}{q^2} \right)^{-1/2} \cdot \sum_{\star \mathscr{F}\in H^{\tilde{p}+1}(M,\mathbb{Z})} e^{-\tilde{I}_\text{cl}(\star \mathscr{F})}\nonumber\\
&= \det \left( \frac{2 \pi \Gamma_{p+1}}{q^2} \right)^{-1/2} \cdot \tilde Z_\text{inst}.
\end{align}
In the second line we used 
\begin{equation}
\tilde q = \frac{2 \pi}{q}
\end{equation}
and Poincar\'{e} duality, which implies $\Gamma_{k} = \Gamma_{D-k}^{-1}.$\footnote{More precisely, duality implies only that $\Gamma_k = \Gamma_{D-k}^{-1}$ up to a matrix that is invertible over the integers (and therefore has unit determinant). We can set this matrix to the identity by a choice of basis.}

Now we compute the ratio of partition functions, i.e. the duality anomaly. Using \eqref{oscillatorz}, \eqref{zeromodez} and \eqref{bundleduality},
\begin{equation}
\label{zratio}
\frac{Z_\text{osc}Z_\text{zero}Z_\text{inst} }{ \tilde{Z}_\text{osc}\tilde{Z}_\text{zero}\tilde{Z}_\text{inst}}
= \frac{\prod_{k=0}^p \left[\det(\Delta_k)^{(-1)^{p+1-k}(p+1-k)/2} \cdot \det\left(\frac{2\pi}{q^2} \Gamma_{p-k}\right)^{(-1)^k/2}\right] \det\left(\frac{2\pi}{q^2} \Gamma_{p+1}\right)^{-1/2} }{\prod_{k=0}^{\tilde{p}} \left[\det(\Delta_k)^{(-1)^{\tilde{p}+1-k}(\tilde{p}+1-k)/2} \cdot \det\left(\frac{2\pi}{\tilde{q}^2} \Gamma_{\tilde{p}-k}\right)^{(-1)^k/2}\right]}.
\end{equation}
First we compute the ratio of oscillator determinants, which was first calculated by \cite{Schwarz1984}. It is convenient to decompose $\Delta_k = \delta_k d_k + d_{k-1} \delta_{k-1}$\footnote{For $k=0$ or $D$, only the nontrivial term contributes.}; the spectra of $\delta_k d_k$ and $d_k \delta_k$ agree up to zero modes, so $\det(\delta_k d_k)=\det(d_k \delta_k)$. Then
\begin{equation}
Z_\text{osc} = \prod_{k=0}^p \det(\delta_k d_k)^{(-1)^{p+1-k}(p+1-k)/2} \ \prod_{k=1}^p \det(d_{k-1} \delta_{k-1})^{(-1)^{p+1-k}(p+1-k)/2}
= \prod_{k=0}^p \det(\delta_k d_k)^{(-1)^{p+1-k}/2}.
\end{equation}
Letting $E_k = \det(\delta_k d_k)$, the ratio of oscillator partition functions is
\begin{align}
\label{oscratio}
\frac{Z_\text{osc}}{\tilde{Z}_\text{osc}} &= \prod_{k=0}^p E_k^{(-1)^{p+1-k}/2}\ \prod_{k=0}^{\tilde{p}} E_k^{(-1)^{\tilde{p}-k}/2}\nonumber\\
&= \prod_{k=0}^p E_k^{(-1)^{p+1-k}/2}\ \prod_{k=0}^{D-p-2} E_{D-k-1}^{(-1)^{D-p-2-k}/2}= \left[\prod_{k=0}^{D-1} E_k^{(-1)^{k-1}/2}\right]^{(-1)^p }.
\end{align}
This expression is a power of the Ray-Singer analytic torsion
\begin{equation}
\label{osctors}
\tau_\text{RS} = \prod_{k=0}^{D-1} E_{D-k-1}^{(-1)^k/2} = \prod_{k=0}^D \left(\det \Delta_k\right)^{k(-1)^{k+1}/2}.
\end{equation}
In even $D$, $\tau_\text{RS}= 1$ by Poincar{\'e} duality ($E_k = E_{D-k-1}$). Comparing \eqref{oscratio} and \eqref{osctors}, the ratio of oscillator partition functions is
\begin{equation}
\label{eq:apdxoscratio}
\log\frac{Z_\text{osc}}{\tilde{Z}_\text{osc}} =
\left\{
	\begin{array}{ll}
		0  &\quad \quad D =2n \\
		(-1)^{p+1}\ \log \tau_\text{RS} & \quad \quad D=2n+1
	\end{array}
	\right.
\end{equation}
which was the result of \cite{Schwarz1984}.

Next up are the zero modes and instantons. From \eqref{zratio} we get
\begin{align}
\label{eq:zminstratio}
\frac{Z_\text{zero} Z_\text{inst}}{\tilde{Z}_\text{zero}\tilde{Z}_\text{inst}} &= \prod_{k=0}^p \det\left(\frac{2\pi}{q^2}\Gamma_k\right)^{(-1)^{p-k}/2} \det\left(\frac{2\pi}{q^2} \Gamma_{p+1}\right)^{-1/2}\prod_{k=0}^{\tilde{p}} \det\left(\frac{\tilde{q}^2}{2\pi} \Gamma_k^{-1}\right)^{(-1)^{\tilde{p}-k}/2}
\nonumber\\
&=  \prod_{k=0}^{p+1} \det\left(\frac{2\pi}{q^2}\Gamma_k\right)^{(-1)^{p-k}/2}\ \prod_{k=0}^{D-p-2} \det\left(\frac{2\pi}{q^2} \Gamma_{D-k}\right)^{(-1)^{D-p-2-k}/2}\nonumber\\
&=\left[ \prod_{k=0}^{D} \left(\frac{2\pi}{q^2}\right)^{(-1)^kb_k}\det\left(\Gamma_k\right)^{(-1)^{k}}\right]^{(-1)^p/2}\nonumber\\
&=\left[ \left(\frac{2\pi}{q^2}\right)^{\chi}\prod_{k=0}^{D} \det\left(\Gamma_k\right)^{(-1)^{k}}\right]^{(-1)^p/2}
\end{align}
where we used Poincar{\'e} duality in the second line, then zeta-function regularization in the third to pull out the charge factors.\footnote{We define our determinants using zeta-function regularization, $\det \Delta = e^{-\zeta'_{\Delta}|_{s=0}}$, and so $\det(a\Delta) = a^{\zeta_{\Delta}|_{s=0}}\det\Delta$ for scalar $a$. The fact that $\zeta_\Delta|_{s=0} = -\mbox{dim ker } \Delta$ \cite{Rosenberg1997} (up to a term in even $D$ that can be absorbed into a local counterterm \cite{Donnelly:2015hxa}) then implies the equality in the third line of \eqref{eq:zminstratio}.} In even $D=2n$, the $\det\Gamma$s cancel pairwise, while the ratio in odd dimensions is a power of the Reidemeister torsion:
\begin{equation}
\tau_\text{Reid} = \prod_{k=0}^D \det \Gamma_k^{(-1)^{k}/2}.
\end{equation}
Rewriting \eqref{eq:zminstratio} as
\begin{equation}
\label{eq:apdxzminstratio}
\frac{Z_\text{zero} Z_\text{inst}}{\tilde{Z}_\text{zero}\tilde{Z}_\text{inst}} =
\left\{
	\begin{array}{ll}
		(-1)^{p+1}\ \chi(M) \log\sqrt{\frac{q}{\tilde{q}}}  &\quad \quad D =2n \\
		(-1)^{p}\ \log \tau_\text{Reid} & \quad \quad D=2n+1,
	\end{array}
	\right.
\end{equation}
and using the Cheeger-M\"uller theorem
\begin{equation}
    \tau_\text{RS} = \tau_\text{Reid},
\end{equation}
the duality anomaly of a $p$-form theory on a $D$-manifold $M$ is
\begin{align}
\label{eq:electroanomaly}
\log \frac{Z_p} {\tilde{Z}_{\tilde{p}}} &=
\left\{
	\begin{array}{ll}
	    (-1)^{p+1}\ \chi(M) \log\sqrt{\frac{q}{\tilde{q}}}  &\quad \quad D=2n \\
		(-1)^{p+1}\log\frac{\tau_\text{RS}}{\tau_\text{Reid}}=0, & \quad \quad D=2n+1.  
	\end{array}
\right.
\end{align}
The quantity $q/\tilde{q}$ has mass dimension $2(p+1)-D$. However, the dimensionful factor $\mu$ in the measure \eqref{measure} renders the argument of the log dimensionless, after proper consideration of an anomaly in zeta-function regularization discussed at the end of this appendix.

Next we consider the possibility that some of the cohomology groups $H^k(M,\mathbb{Z})$ have nontrivial torsion subgroups.\footnote{We apologize to the reader for introducing a third kind of torsion.} The partition functions for this more general case were explained in \S~\ref{section:partitionfunctions}, see \eqref{ztors} and \eqref{bundlez}. Separating out the contribution of these factors as
\begin{equation}
Z_p = Z_{p,\text{tors}} Z_{p,\text{free}}
\end{equation}
where $Z_{p,\text{tors}}$ contains the contribution of all the torsion subgroups, we have
\begin{equation}
\label{eq:tors}
Z_{p,\text{tors}} = \prod_{k=0}^{p+1} |T^{p+1-k}|^{(-1)^k}
\end{equation}
where $|T^k|$ is the size of the torsion subgroup of $H^k(M,\mathbb{Z})$.

We need the duality relation for torsion subgroups to proceed further. It is common mathematical knowledge \cite{hatcherp194} that $H^k(M,\mathbb{Z})\cong (H_k/T_k)\oplus T_{k-1}$ and so $T^k=T_{k-1}$. Together with Poincar\'{e} duality $(H^k\cong H_{D-k})$ this implies $T^k \cong T^{D-k+1}$. Using \eqref{eq:tors}, we find
\begin{align}
\label{eq:torsratio}
\frac{Z_{p,\text{tors}}}{\tilde{Z}_{\tilde{p},\text{tors}}} &= \frac{\prod_{k=0}^{p+1} |T^{p+1-k}|^{(-1)^k}}{\prod_{k=0}^{\tilde{p}+1} |T^{\tilde{p}+1-k}|^{(-1)^k}}\nonumber\\
&= \prod_{k=0}^{p+1} |T^{p+1-k}|^{(-1)^k} \prod_{k=0}^{D-p-1} |T^{p+2+k}|^{(-1)^{k+1}}\nonumber\\
&= \left[ \prod_{k=0}^D |T^k|^{(-1)^{k+1}}\right]^{(-1)^p}
\end{align}
which is equal to 1 by Poincar\'{e} duality in even $D$. The odd-$D$ duality relation becomes
\begin{equation}
\frac{Z_p}{\tilde{Z}_{\tilde{p}}} = \left(\frac{\tau_\text{RS}}{\tau_\text{Reid}}\right)^{(-1)^{p+1}} \left[\prod_{k=0}^D |T^k|^{(-1)^{k+1}}\right]^{(-1)^p}.
\end{equation}
Thus it appears at first as if we have recovered a nontrivial anomaly. However, Cheeger \cite{Cheeger1979} showed that the relation between Ray-Singer and Reidemeister torsion is modified in the presence of torsion subgroups\footnote{We found appendix E of \cite{Metlitski:2015yqa} particularly helpful in understanding these results.}. The general relation is

\begin{equation}
\frac{\tau_\text{RS}}{\tau_\text{Reid}}=\prod_{k=0}^D |T^k|^{(-1)^{k+1}}
\end{equation}
and so the odd-dimensional duality remains trivial even on manifolds whose integral cohomologies have nontrivial torsion subgroups. This completes our proof that the odd-dimensional duality anomaly vanishes.

Last we discuss the measure factors in \eqref{measure} which correct the units in \eqref{eq:electroanomaly}. We define our functional determinants using zeta-function regularization, $\det \Delta = e^{-\zeta'_{\Delta}|_{s=0}}$, and so $\det(a\Delta) = a^{\zeta_{\Delta}|_{s=0}}\det\Delta$ for scalar $a$. In even dimensions there is an anomaly\footnote{We apologize to the reader for introducing a third kind of anomaly.} in $\zeta_\Delta|_{s=0}$: it is given by \cite{Rosenberg1997}
\begin{equation}
\label{scaleanom}
\zeta_{\Delta_k}|_{s=0} = \mathcal{A}_k -\text{dim ker }\Delta_k
\end{equation}
where $\mathcal{A}_k=a^{(k)}_{D/2}$ is the coefficient of $t^{0}$ in the asymptotic expansion of the trace of the heat kernel $\tr e^{-t \Delta_k}$ as $t \to 0$. 
They satisfy a Betti number-like relation to the Euler character, the McKean-Singer formula \cite{McKean1967}:
\begin{equation}
\label{mcks}
\sum_{k=0}^D (-1)^k \mathcal{A}_k = \chi(M).
\end{equation}
In odd dimensions $\mathcal{A}_k=0$ for all $k$.

The full partition with all the factors of $\mu$ was given in \eqref{eq:partitionfunction}:
\begin{equation}
Z_p = \prod_{k=0}^{p} \left( \frac{\det \left( \frac{2 \pi\mu^{2(p-k+1)}}{q^2} \Gamma_k \right)^{1/2} }{ |T^k| \det\left(\frac{\Delta_k}{\mu^2}\right)^{(p-k+1)/2} } \right)^{(-1)^{p-k}}
|T^{p+1}| \sum_{\mathscr{F} \in \mathbb{Z}^{b_{p+1}}} e^{-I[\mathscr{F}]}.
\end{equation}
Scaling out the factors of $\mu$ in the functional determinants using \eqref{scaleanom}, it is obvious that the nonanomalous pieces in the rescaling cancel with the factors of $\mu$ from the zero modes. However, we must deal with the anomalous piece. We find
\begin{align}
Z_\text{osc} Z_\text{zero} = \prod_{k=0}^{p} \mu^{-\mathcal{A}_k(p+1-k)(-1)^{p+1-k}} \left( \frac{\det \left( \frac{2 \pi}{q^2} \Gamma_k \right)^{1/2} }{ |T^p| \det(\Delta_k)^{(p-k+1)/2} } \right)^{(-1)^{p-k}}.
\end{align}
Note that we have defined the dimensionful measure factors for the ghosts to be given by the same $\mu$ as appeared in the measure for the $p$-forms; this is part of our definition of the theory. We will also take the dual measure factors $\tilde\mu$ equal to the original $\mu$. With these choices, the net effect of the anomaly is to multiply the ratio of partition functions \eqref{zratio} by
\begin{equation}
\label{scaleanomfactor}
\prod_{k=0}^{p} \mu^{-\mathcal{A}_k(p+1-k)(-1)^{p+1-k}}\prod_{k=0}^{\tilde{p}} \mu^{\mathcal{A}_k(\tilde{p}+1-k)(-1)^{\tilde{p}+1-k}}.
\end{equation}
The log of \eqref{zratio} picks up a piece
\begin{align}
\label{scaleanomlog}
\sum_{k=0}^{D}(-1)^{p-k}\mathcal{A}_k (p+1-k)&=(-1)^p\left[(p+1)\sum_{k=0}^{D}(-1)^{k}\mathcal{A}_k - \sum_{k=0}^{D}(-1)^{k}\frac{k}{2} \mathcal{A}_k- \sum_{k=0}^{D}(-1)^{k}\frac{(D-k)}{2}\mathcal{A}_k\right]\nonumber\\
&= (-1)^p \left(p+1-\frac{D}{2}\right)\chi(M)
\end{align}
multiplied by $\log\mu$. The LHS of \eqref{scaleanomlog} follows from \eqref{scaleanomfactor} after using Poincar\'e duality, $\mathcal{A}_k = \mathcal{A}_{D-k}$. The first equality follows from Poincar\'e duality, the second from \eqref{mcks}. Thus the net effect of the measure factors is to correct the even-dimensional duality relation to 
\begin{equation}
\log \frac{Z_p} {\tilde{Z}_{\tilde{p}}} =(-1)^{p+1}\ \chi(M) \log\sqrt{\frac{q/\tilde{q}}{\mu^{2(p+1)-D}}},
\end{equation}
which is dimensionless as expected.

\section{With a $\theta$ term}
\label{appendix:thetaterm}

In this appendix we extend our analysis of the duality anomaly of $p$-form Maxwell theories to include a topological term. We focus on terms of the form
\begin{equation}
\theta \int_M F\wedge F
\end{equation}
which only exist when $p+1 = D/2$ and vanish unless $D$ is a multiple of 4. For concreteness we describe the case of $p=1$ duality with a $\theta$ term, which was studied by \cite{Witten1995} (and earlier \cite{Montonen:1977sn}); our analysis extends to all $D=4n$.

We start with the action
\begin{equation}
I=\frac{2\pi^2}{q^2}\left(\frac{1}{8\pi^2}\int_M F_{\mu\nu}F^{\mu\nu}\right) + \frac{i\theta}{2}\left(\frac{1}{8\pi^2}\int_M \frac{1}{2} \epsilon_{\mu\nu\rho\sigma}F^{\mu\nu}F^{\rho\sigma}\right).
\end{equation}
The terms in parentheses are the metric on the middle cohomology and the intersection form, respectively. The intersection form calculates the winding number of $F$ and depends only on its cohomology. On a spin manifold, the intersection form is even, and the action is invariant under modular transformations.\footnote{If the manifold does not admit a spin structure, the invariance is under the Hecke group generated by S and the T-transformation $\tau\rightarrow \tau+2$ \cite{Olive2000}.}

The $\theta$ term only affects the instanton contribution to the partition function. As described in \S~\ref{section:partitionfunctions}, the instanton contribution to the action is
\footnote{
In this section $\Gamma$ refers to $\Gamma_2$.}
\begin{align}
I_\text{inst}&= \frac{2\pi^2}{q^2}\vec{m}\cdot \Gamma \cdot\vec{m} + \frac{i\theta}{2} \vec{m}\cdot P\cdot \vec{m}\nonumber\\
&= \vec{m}\cdot \Gamma \left(\frac{2\pi^2}{q^2}+\frac{i\theta}{2}S\right)\cdot\vec{m}\nonumber\\
&= i\pi(\bar{\tau}\ \vec{m}_+ \cdot \Gamma_+ \cdot \vec{m}_+ -\tau\ \vec{m}_- \cdot \Gamma_- \cdot \vec{m}_-)
\end{align}
In the first line we defined the matrix $P_{ij} = \int w_i \wedge w_j$. In the second line we related $P$ to $\Gamma$ via $P=\Gamma S$, where $S$ is defined by $\star w_i = S_{ij} w_j$. This notation follows \cite{Donnelly:2013tia}. In the last line we used the orthogonal decomposition of the middle homology into self-dual and anti-self-dual forms, writing $\vec{m}=(\vec{m}_+, \vec{m}_-)$, where $\vec{m}_{\pm}$ are basis vectors for $H^2_{\pm}(M,\mathbb{Z})$ and have $b_2^{\pm}$ components each. In this basis $S$ takes the form diag(1,-1).

Now we can get cracking. Using Poisson summation, the instanton partition function is
\begin{align}
\label{inst_theta}
Z_\text{inst} &= \sum_{\vec{m}\in H^2(M,\mathbb{Z})} e^{-\vec{m}\cdot \Gamma\left(\frac{2\pi^2}{q^2}+\frac{i\theta}{2} S\right)\cdot\vec{m}}\nonumber\\
&=\det\left(\frac{\pi}{\Gamma(2\pi^2/q^2+i\theta S/2)}\right)^{1/2} \sum_{\vec{n}\in H_2(M,\mathbb{Z})} e^{- \vec{n}\cdot \Gamma^{-1}\left(\frac{2\pi^2}{q^2}+\frac{i\theta}{2}S\right)^{-1}\cdot \vec{n} \cdot \pi^2}
\end{align}
Consider the determinant out front first:
\begin{align}
\label{poisson_det}
\det\left(\frac{\pi}{\Gamma(2\pi^2/q^2+i\theta S/2)}\right)^{1/2}&= \det\Gamma^{-1/2}\ \mbox{det}^+ \left(\frac{2\pi^2}{q^2}+\frac{i\theta S}{2}\right)^{-1/2}\mbox{det}^- \left(\frac{2\pi^2}{q^2}+\frac{i\theta S}{2}\right)^{-1/2}\nonumber\\
&= \det\Gamma^{-1/2} \ \left(\frac{2\pi^2}{q^2}+\frac{i\theta }{2}\right)^{-b_2^+/2}
\left(\frac{2\pi^2}{q^2}-\frac{i\theta }{2}\right)^{-b_2^-/2}\nonumber\\
&= \det \Gamma^{-1/2} (i\bar{\tau})^{-b_2^+/2}(-i\tau)^{-b_2^-/2}.
\end{align}
In the first line we used the decomposition $H^2 = H^2_+\oplus H^2_-$. 

Next, the exponent of \eqref{inst_theta}, which is
\begin{align}
\pi^2 \vec{n}\cdot \Gamma^{-1}\left(\frac{2\pi^2}{q^2}+\frac{i\theta}{2}S\right)^{-1}\cdot \vec{n}&= \pi (\vec{n}_+\ \vec{n}_-) \Gamma^{-1} \left( \begin{array}{cc}
\frac{2\pi}{q^2}+\frac{i\theta}{2\pi} & 0 \\
0 & \frac{2\pi}{q^2}-\frac{i\theta}{2\pi} \\
\end{array} \right)^{-1}\left( \begin{array}{c}
\vec{n}_+ \\
\vec{n}_- \\
\end{array} \right)\nonumber\\
&= i\pi \left[\left(\frac{-1}{\bar{\tau}}\right) \vec{n}_+\cdot \Gamma_+ \cdot \vec{n}_+ - \left(\frac{-1}{\tau}\right) \vec{n}_- \cdot \Gamma_-\cdot  \vec{n}_-\right]\\
&= \tilde{I}_\text{inst}(\star \mathscr{F})
\end{align}
i.e. the action of an instanton in the dual theory. The full $p$-form partition function is
\begin{align}
Z_p &= Z_\text{osc} Z_\text{osc} Z_\text{inst}\nonumber\\
&= Z_\text{osc} \tilde{Z}_\text{inst}\det\left(\frac{2\pi}{q^2}\Gamma_{0}\right)^{-1/2}\det\left(\frac{2\pi}{q^2}\Gamma_{1}\right)^{1/2}
\det \Gamma_2^{-1/2} (i\bar{\tau})^{-b_2^+/2}(-i\tau)^{-b_2^-/2} \nonumber\\
&= Z_\text{osc} \tilde{Z}_\text{inst} \det \Gamma_0^{-1/2}\det \Gamma_1^{1/2} \det \Gamma_2^{-1/2} \left(\frac{2\pi}{q^2}\right)^{(b_1-b_0)/2}(i\bar{\tau})^{-b_2^+/2}(-i\tau)^{-b_2^-/2}\nonumber\\
&= \tilde{Z}_\text{osc}\tilde{Z}_\text{inst}\det \Gamma_0^{-1/2}\det \Gamma_1^{1/2} \det \Gamma_2^{-1/2} \left(\text{Im } \tau\right)^{(b_1-b_0)/2} (i\bar{\tau})^{-b_2^+/2}(-i\tau)^{-b_2^-/2}
\end{align}
while the dual partition function is
\begin{align}
\tilde{Z}_{\tilde{p}} &= \tilde{Z}_\text{osc} \tilde{Z}_\text{zero} \tilde{Z}_\text{inst}\nonumber\\
&= \tilde{Z}_\text{osc}  \tilde{Z}_\text{inst}\det\left(\frac{2\pi}{\tilde{q}^2}\Gamma_{0}\right)^{-1/2}\det\left(\frac{2\pi}{\tilde{q}^2}\Gamma_{1}\right)^{1/2}
\nonumber\\
&= \tilde{Z}_\text{osc}\tilde{Z}_\text{inst}\det \Gamma_0^{-1/2}\det \Gamma_1^{1/2}   \left(\text{Im }\frac{-1}{\tau}\right)^{(b_1-b_0)/2}.
\end{align}
It then follows from $\text{Im }\frac{-1}{\tau} = \frac{\text{Im }\tau}{\tau\bar{\tau}}$ and $\det\Gamma_2=1$ that\footnote{$\Gamma_2$ is a positive matrix, and $P^{-1}$ is an integer matrix (by Poincar\'{e} duality), so $|\det\Gamma_2|=1$.}
\begin{align}
\label{thetaduality}
\tilde{Z} &= Z\left(\frac{-1}{\tau}\right)= e^{i\pi\sigma/4} \bar{\tau}^{(b_2^+-b_1+b_0)/2}\tau^{(b_2^- -b_1+b_0)/2}Z(\tau)\nonumber\\
&= e^{i\pi\sigma/4} \bar{\tau}^{(\chi+\sigma)/4}\tau^{(\chi-\sigma)/4}Z(\tau).
\end{align}
From eq. \eqref{thetaduality} we see that the partition function transforms as a modular form, up to a phase. This agrees with Witten's result for the duality anomaly \cite{Witten1995} except for the phase, which also appears in the calculation of the anomaly by Alvarez and Olive in \cite{Olive2000}, and in Metlitski's calculation in \cite{Metlitski:2015yqa}. It seems that the partition function is a modular form only up to this phase, whose presence allowed us to make a physical argument about the topological signature of the replica manifold in \S~\ref{subsection:thetaterm}.

While $\theta\int F\wedge F$ vanishes in 0-form theory in two dimensions, in 1-form theory there is a term $\theta \int F$ which plays a similar role to the $\theta$ term in $D=4n$ dimensions (i.e. it assigns a phase linear in $n$ to instantons in the path integral). However, since it is linear in the field strength, it enters differently into the duality relation. The action
\begin{equation}
S = \frac{1}{2}\int F\wedge \star F + \frac{i\theta q}{2\pi}\int F
\end{equation}
leads to the bundle sum
\begin{align}
Z_\text{inst} &= \sum_{m\in \mathbb{Z}} e^{-\frac12\left(\frac{2\pi}{q}\right)^2 m^2+ i\theta m}\nonumber\\
&=\det\left(\frac{2\pi}{q^2}\Gamma_2\right)^{-1/2} \sum_{n\in \mathbb{Z}} e^{-\frac{1}{2}\left(\frac{2\pi}{\tilde{q}}\right)^2\left(n+\frac{\theta}{2\pi}\right)^2}.
\end{align}
The shift by $\theta$ in the lattice of integer charges in the dual theory is the Witten effect \cite{Witten1979}. 
However, since it does not enter into the determinant, $\theta$ does not affect the duality anomaly in $D=2$.

\section{One-dimensional Maxwell theory}
\label{appendix:1dmaxwell}

Maxwell theory in one spacetime dimension is a particularly trivial theory. It has only a vacuum state, so the canonical partition function is simply
\begin{equation}
Z_\text{canonical} = \sum_{\text{states}} e^{-\beta E} = 1.
\end{equation}
We can use the sledgehammer forged in the body of this work to reproduce this trivial result. 
This exercise serves to clarify the role of the non-oscillator contributions in a simple example.

The oscillator partition function is 
\begin{equation}
Z_\text{osc} = \left(\det \Delta_1\right)^{-1/2}\left[\left(\det \Delta_0\right)^{1/2}\right]^2 =  \left(\det \Delta_0\right)^{1/2}
\end{equation}
where the last equality follows from Poinar\'{e} duality. As usual, we will calculate the functional determinant using $\zeta$-function regularization: $\det \Delta = e^{-\zeta_{\Delta}'(0)}$. Noting that the eigenfunctions $f_n$ of the scalar Laplacian take the form $f_n(\theta)\sim e^{\frac{2\pi i n \theta}{\beta}}$, the relevant zeta function reads
\begin{equation}
\zeta_{\Delta}(s) = \sum_{n\neq 0} \frac{1}{\lambda_n^s} = 2\left[-\left(\frac{\beta}{2\pi}\right)^2\right] \zeta_R(2s)
\end{equation}
where in the second equality we introduced the Riemann zeta function $\zeta_R(s) = \sum_{n\in\mathbb{Z}^+} \frac{1}{n^s}$.

Evaluation then yields
\begin{equation}
\zeta_{\Delta}'(0) = 4\zeta_R'(0) + 2\log\left[-\left(\frac{\beta}{2\pi}\right)^2\right]\zeta_R(0)= -\beta^2
\end{equation}
and so 
\begin{equation}
Z_\text{osc} = \beta \neq Z_\text{canonical}.
\end{equation}
However, the zero mode partition function is
\begin{align}
Z_\text{zero} &= \det\left(\frac{2\pi}{q^2}\Gamma_1\right)^{1/2} \det\left(\frac{2\pi}{q^2}\Gamma_0\right)^{-1/2}\nonumber\\
&= \det \Gamma_0^{-1} = \beta^{-1}
\end{align}
where in the last line we used Poincar\'{e} duality. There are no instantons in this simple scenario so we are done. We now see that
\begin{equation}
Z_\text{osc} \cdot Z_\text{zero} = 1 = Z_\text{canonical},
\end{equation}
i.e. the gauge theory partition function agrees with the canonical partition function only once the functional determinants are combined with the contribution from the zero modes.

\bibliographystyle{utphys}
\bibliography{duality}

\end{document}